\documentclass{article}
\usepackage{amssymb}
\usepackage{amsmath}

\begin{document}

\begin{center}
{\large\bf Classical and Quantum-Mechanical Axioms with the Higher Time Derivative
Formalism } \bigskip 

Timur Kamalov

Moscow State Open University, P. Korchagina street, 22, Moscow,

107996, Russia

Moscow Polytechnic Institute in Moscow State University of Mechanical
Engineering, P. Korchagina street, 22, Moscow, 107996, Russia

E-mail: timkamalov@gmail.com

Home Page: http://www.timkamalov.narod.ru/
\end{center}

\textit{A Newtonian mechanics model is essentially the model of a point body
in an inertial reference frame. How to describe extended bodies in
non-inertial (vibration) reference frames with the random initial
conditions? One of the most generalized ways of descriptions (known as the
higher derivatives formalism) consists in taking into account the infinite
number of the higher temporal derivatives of the coordinates in the Lagrange
function. Such formalism describing physical objects in the infinite
dimensions space does not contradict to the quantum mechanics and infinite
dimensions Hilbert space}

\section{Introduction}

Classical Newtonian mechanics is essentially the simplest way of mechanical
system description with second-order differential equations, when higher
order time derivatives of coordinates can be neglected. The extended model
of Newtonian mechanics with higher time derivatives of coordinates is based
on generalization of Newton's Laws onto arbitrary reference systems (both
inertial and non-inertial ones) with body dynamics being described with
higher order differential equations. Newton's Laws, constituting, from the
mathematical viewpoint, the axiomatic of classical physics, actually
postulate the assertion that the equations describing the dynamics of bodies
in inertial reference frames are second-order differential equations.
However, the actual time-space is almost without exception non-inertial, as
it is almost without exception that there exist (at least weak) fields,
waves, or forces perturbing an ideal inertial reference frames. It
corresponds to Mach's principle [1] with a general statement "Local physical
laws are determined by the large-scale structure of the Universe."
Non-inertial nature of the actual time-space is also supported by
observations of the practical astronomy that expansion of the Universe
occurs with an acceleration. In other words, actually any real reference
frame is a non-inertial one, and such physical reality can be described by a
differential equation with time derivatives of coordinates of an order
exceeding two, which play the role of additional variables. Aristotle's
physics considered velocity to be proportional to the applied force, hence
the body dynamics is described by a first-derivative differential equation.
Classical Physics in the inertial reference frames describe a free body
maintains the constant velocity of the translational motion. In this case,
the body dynamics is described by a second-order differential equation, with
acceleration being proportional to the force [2]. This corresponds to the
Lagrange function depending on coordinates and their first derivatives
(velocities) of the body, and the Euler--Lagrange equation resulting from
the principle of the least action. This model of the physical reality
describes macrocosm fairly well, but it fails to describe microparticles.
Both Newtonian axiomatic and the second Newton law are not valid in the
microcosm. Only averaged values of observable physical quantities yield in
the microcosm an approximate analogue of the second Newton law; this is the
so-called Ehrenfest's theorem. The Ehrenfest's equation yields an averaged,
rather than precise, ratio between the second time derivative of coordinate
and the force, while to describe the scatter of quantum observables, the
probability theory apparatus is required. Since Newtonian dynamics is
restricted to the second-order derivatives, while microobjects must be
described by equations with additional variables, making the Planck's
constant tend to zero corresponds to neglecting these variables. Hence,
offering the model of extended Newtonian dynamics, we consider the classical
and quantum theories with additional variables, describing the body dynamics
with higher-order differential equations. In our model, the Lagrangian is
considered to be dependent not only on coordinates and their first time
derivatives, but also on higher-order time derivatives of coordinates.
Classical dynamics of a test particle's motion with higher-order time
derivatives of the coordinates was first described in 1850 by Ostrogradski
[3] and is known as Ostrogradski's Formalism. Being a mathematician,
Ostrogradski considered the coordinate systems rather than the reference
frames. This is just the case corresponding to a real reference frame
comprising both inertial and non-inertial reference frames. In the general
case, the Lagrangian takes on the form
\begin{equation}
L=L(q,\dot{q},\ddot{q},...,\dot{q}^{(n)},...).  \label{1}
\end{equation}

Attempts to build a unified theory for both quantum and classical mechanics,
exemplified by the Hybrid theory of classical and quantum mechanics [4], are
natural and make sense. However, constructing a theory without an axiomatic
consistency with both theories resembles a construction without foundation.
The point is that the systems of axioms of classical and quantum theories
are mutually incompatible and even contradictory. Obviously, the
mathematical constructions of united theory may prove questionable without
presenting a framework which is conceptually consistent both with classical
and quantum mechanics. For example, a natural question arises: "Can the
phase space be used, and can the momentum and the coordinate exist
simultaneously in quantum mechanics?" The Heisenberg uncertainty principle
tells us that it is not possible.

At the same time, quantum theory describes objects in Hilbert space, i.e. in
terms of an infinite number of variables, and thus it gives a more detailed
description as compared to classical theory. Thus, the question of
quantum-mechanical description being incomplete should be answered so that
any theory as a model of physical reality is incomplete following Godel's
theorem.

\qquad The classical description of physical reality contains an
incomparably fewer number of variables. This raises the question: "How the
classical description can be completed?" While a possibility of
supplementing the quantum mechanical description with additional ("hidden")
variables has been debated for long, the question as to how to complete the
classical description to make it compatible with the quantum mechanical one
has not received a due attention.

The classical and quantum theory could be based on the following common
axioms:

1. Any reference frame is subject to random external influences. Hence,
every reference frame is individual and a transition from one to another
reference frame may lead to jump like changes. The notion of the inertial
frame in classical mechanics is valid only on the average and,
correspondingly, the Galilean relativity is an average notion as well. Then
there are many trajectories of a particle corresponding to different
reference frames; the Heisenberg uncertainty can be understood as a
consequence of the nonexistence of ideal inertial frames; and the Ehrenfest
theorem can be seen as a consequence of the inertial frame being an average
notion. Correspondingly, \textit{the free body preserves the same order of
its time derivative as the constant kinematic characteristics of the class
of reference frames}, e.g. in the uniformly accelerating reference frame the
free body preserves its acceleration.

2. In Hybrid theory [4] the generalized Ehrenfest relations for the QM
observables are defined by coordinates average. Compare the averaging
procedure in this paper with the time averaging. Within the above framework
were the ideal inertial frames are non-existent, we can consider the
averaging of the classical equations of motion over the time interval $%
\Delta t$:

\begin{center}
$-\frac{\partial U}{\partial q}=\frac{d}{dt}\frac{[p(t+\Delta t)+p(t-\Delta
t)]}{2}.$
\end{center}

Using the Taylor expansion

\begin{center}
$p(t\pm \Delta t)=p(t)\pm \overset{\cdot }{p}\Delta t+\frac{1}{2!}\overset{%
\cdot \cdot }{p}(t)\Delta t^{2}+...+\frac{(-1)^{n}}{n!}\overset{\cdot }{p}%
^{(n)}(t)\Delta t^{n}+..$
\end{center}

the function $F=-\frac{\partial U}{\partial q}$ can be expanded as follows:

\begin{center}
$F(q.\overset{\cdot }{q},\overset{\cdot \cdot }{q},\overset{\cdot \cdot
\cdot }{q}...,\overset{\cdot }{q}^{(k)})=\overset{\cdot }{p}(t)+\frac{1}{2!}%
\overset{\cdot }{p}^{(3)}(t)\Delta t^{2}+\frac{1}{4!}\overset{\cdot }{p}%
^{(5)}(t)\Delta t^{4}+...,$
\end{center}

where $\overset{\cdot }{p}^{(n)}$ denotes $n$-th time derivative of momentum
$p$. \textit{It is the Extended Law of Dynamics in arbitrary reference frames%
} including the case of the vibration non-inertial reference frames.
Correspondingly, the free body preserves the same order of its time
derivative like the constant kinematic characteristics of the reference
frames. For example, in the uniformly accelerating reference frame the free
body preserves its acceleration.

3. The de-Broglie waves $\psi =\psi _{0}\exp (-iS/\hbar )$\ with the actions
functions $S=S(q,\dot{q},\ddot{q},...,\dot{q}^{(n)},...)$\ can be considered
as having the gravity-inertial nature following from the fact that every
\textit{reference frame is vibrational due to the influence of random
gravitational fields and waves} so that every free particle appears to be
oscillating.

4. As the action function $S=S(q,\dot{q},\ddot{q},...,\dot{q}^{(n)},...)$ is
a convergent series in high derivatives of $q$ the difference $\left\vert
S(q,\dot{q},\ddot{q},...,\dot{q}^{(n)},...)-S(q,\dot{q})\right\vert =h$ is
finite and can be identified with the constant $h$. Within the presented
framework the variables of \textit{the (high order extension of the) phase
space do describe the completed dynamics of a particle}, but they cannot be
measured because the ideal inertial reference frames do not exist in really.
The infinite dimensionality of Hilbert space can also be understood as a
consequence of all high order time derivatives being taken into account in
the description of the dynamics.

\section{How to Complete the Quantum-Mechanical Description?}

If the statement by Einstein, Podolsky and Rosen on incompleteness of
Quantum-Mechanical description of nature is correct, then we can regard
Quantum Mechanics as a Method of Indirect Computation. The problem is,
whether the theory is incomplete or the nature itself does not allow
complete description? And if the first option is correct, how is it possible
to complete the Quantum-Mechanical description? Here we try to complement
de-Broglie's idea on wave-pilot the stochastic inertial-gravitation gives
origin to. We assume that de-Broglie's wave-pilots are
inertial-gravitational stochastic ones with the high derivatives, and we
shall regard micro-objects as test classical particles being subject to the
influence of de-Broglie's waves stochastic inertial-gravitation. The Quantum
Theory exists for many decades. But is everything OK with it
completeness[5]? To our opinion, it is not just so. The incompleteness of
Quantum-Mechanical description gives rise to various paradoxes, such as
Einstein-Podolsky-Rosen (EPR) one, the paradox of the Schrodinger's cat, the
Paradox of Quantum Non-locality and Paradox of the Quantum Teleportation. In
this study we shall call the phenomena of quantum nonlocal behavior and
teleportation of the quantum states as paradoxes because they follow from
Stochastic Gravitation Model of Quantum Mechanic. It can be easily seen that
these are paradoxes, and indeed they are brought about by the drawbacks in
the Quantum Theory rather than being actual properties of nature. This is
due to the fact that time in Quantum Theory plays the role not conforming to
the physical reality. In particular, the Quantum Theory employs the concept
of Hilbert Space, in which time acts as a parameter. Henceforth, this
parameter (i.e. the time) may be the same in different points of the Hilbert
Space. This property of time in the Hilbert Space brings about the effects
of simultaneous quantum states of microobjects at different space points (or
transfer of the state from one Hilbert Space point to another with
velocities exceeding the velocity of light). These effects of the Quantum
Theory that are apparently real we call here the Quantum Non-locality
Paradox. The Paradox of Quantum Teleportation is a sort of Quantum
Non-locality Paradox. These paradoxes do not exist in the Classical Physics
and in the Stochastic Gravitation Model of the Quantum Mechanic, and General
Relativity Theory (employing the 4-dimensional space), in which different
points of time-space correspond to different values of the time. And another
question is whether the quantum-mechanical wave-function interpretation of
micro-objects is complete? Let us select harmonic coordinates (the condition
of harmonicas of coordinates mean selection of concomitant frame $\frac{%
\partial h_{\nu }^{\mu }}{\partial x^{\mu }}=\frac{1}{2}\frac{\partial
h_{\mu }^{\mu }}{\partial x^{\nu }}$) and let us take into consideration
that $h_{\mu \nu }$ satisfies the gravitational field equations
\begin{equation}
\square h_{\mu \nu }(j)=-16\pi GS_{\mu \nu }(j),
\end{equation}

which follow from the General Theory of Relativity; here $S_{\mu \nu }$ is
energy-momentum tensor of gravitational field sources with d'Alemberian $%
\square $ and gravity constant $G$. Then, the solution shall acquire the
form
\begin{equation}
h_{\mu \nu }(j)=e_{\mu \nu }(j)\exp (ik_{\gamma }(j)x^{\gamma })+e_{\mu \nu
}^{\ast }(j)\exp (-ik_{\gamma }(j)x^{\gamma }),
\end{equation}

where the value $h_{\mu \nu }(j)$\ is called metric perturbation, $e_{\mu
\nu }(j)$\ polarization, and $k_{\gamma }(j)$\ is 4-dimensional wave vector.

We shall assume that this metric perturbation $h_{\mu \nu }(j)$ is
distributed in space with an unknown distribution function $\rho =\rho
(h_{\mu \nu })$. Relative oscillations $\ell $ of two particles in classic
gravitational fields are described in the General Theory of Relativity by
deviation equations, which we can write for the stochastic case as
\begin{equation}
\frac{D^{2}}{D\tau ^{2}}\ell ^{\mu }(j)+R_{\nu \alpha \beta }^{\mu }(j)\ell
^{\alpha }\frac{dx^{\nu }}{d\tau }\frac{dx^{\beta }}{d\tau }=F(j),
\end{equation}

being $R_{\nu \alpha \beta }^{\mu }(j)$ the gravitational field Riemann's
tensor with gravitational field number $j$ of the stochastic gravitational
fields and $F(j)$ is the stochastic constant (for the non-stochastic case
this constant is zero $F(j)=0$).

Specifically, the deviation equations give the equations for two particles
oscillations
\begin{equation}
\overset{..}{\ell }^{1}+c^{2}R_{010}^{1}\ell ^{1}=0,\quad \omega =c\sqrt{%
R_{010}^{1}}.
\end{equation}

The solution of this equation has the form
\begin{equation}
\ell ^{1}(j)=\ell _{0}\exp (k_{a}x^{a}+i\omega (j)t),
\end{equation}

being $a=1,2,3$. Each gravitational field or wave with index $j$ and
Riemann's tensor $R_{\nu \alpha \beta }^{\mu }(j)$ shall be corresponding to
the value $\ell ^{\mu }(j)$ with stochastically modulated phase $\Phi
(j)=\omega (j)t$. If we to sum the all fields, we can write the stochastic
phase $\Phi (t)=\omega (t)t$, where $t$ is the time coordinate.

\section{Corrected Bell's Inequalities in Random Gravity-Inertial Fields}

We shall consider the physical model with the Stochastic
Inertial-Gravitational Background [i.e. with the background of
inertial-gravitational fields and waves]. This means that we assume the
non-inertial vibration reference frames due to existence of fluctuations in
inertial-gravitational waves and fields expressed mathematically by metric
fluctuations.

Describing entangled photons in the stochastic curved space, we shall take
into consideration the fact that the scalar product of two 4-vectors $A^{\mu
}$ and $B^{\nu }$ equals $g_{\mu \nu }A^{\mu }B^{\nu }$, where for weak
inertial-gravitational fields one can use the value $h_{\mu \nu }$, which is
the solution of Einstein's equations for the case of weak
inertial-gravitational field in harmonic coordinates.

Correlation factor $M$\ of random variables $\lambda ^{i}$\ [6] which
correspondent to higher-order time derivatives of coordinates are
projections onto directions $A^{\nu }$\ and $B^{n}$\ defined by polarizers
(all these vectors being unit) is
\begin{equation}
\left\vert M_{AB}\right\vert =\left\vert \left\langle AB\right\rangle
\right\vert =\left\vert \langle \lambda ^{\alpha }A^{\beta }g_{\alpha \beta
}\lambda ^{\mu }B^{\nu }g_{\mu \nu }\rangle \right\vert .
\end{equation}

The differential geometry gives

\begin{center}
$\cos \phi =\frac{g_{\alpha \beta }\lambda ^{\alpha }A^{\beta }}{\sqrt{%
\lambda ^{\alpha }\lambda _{\alpha }}\sqrt{A^{\beta }A_{\beta }}}$,

$\cos (\phi +\theta )=\frac{g_{\mu \nu }\lambda ^{\mu }B^{\nu }}{\sqrt{%
\lambda ^{\mu }\lambda _{\mu }}\sqrt{B^{\nu }B_{\nu }}}$.
\end{center}

Here $\alpha ,\beta ,\mu ,\nu $ takes values 0,1,2,3; $\theta $ is angle
between polarizers, then
\begin{equation}
\left\vert M_{AB}\right\vert =\left\vert \frac{1}{\pi }\int_{0}^{\pi }\rho
(\phi )\cos \phi \cos \left( \phi +\theta \right) d\phi \right\vert
=\left\vert \cos \theta \right\vert ,
\end{equation}

here the integral of the distribution function of the metric $\rho $ is $%
\left\vert \frac{1}{\pi }\int_{0}^{\pi }\rho (\phi )d\phi \right\vert =1.$

Finally, the real part of the correlation factor is

\begin{center}
$\left\vert M_{AB}\right\vert =\left\vert \cos \theta \right\vert $.
\end{center}

Then, we obtain the maximum value the Bell's observable $S$ in Rieman's
space for $\theta =\frac{\pi }{4}$

\begin{center}
$\left\vert <S>\right\vert =\left\vert [\langle M_{AB}\rangle +\left\langle
M_{A^{\prime }B}\right\rangle +\left\langle M_{AB^{\prime }}\right\rangle
-\left\langle M_{A^{\prime }B^{\prime }}\right\rangle ]\right\vert =$

$=\left\vert [\cos (-\frac{\pi }{4})+\cos (\frac{\pi }{4})+\cos (\frac{\pi }{%
4})-\cos (\frac{3\pi }{4})]\right\vert =\left\vert \sqrt{2}\right\vert $,
\end{center}

which agrees fairly with the experimental data. The Bell in equality in
Rieman's space shall take on the form ${\left\vert \left\langle
S\right\rangle \right\vert \leq \sqrt{2}}$.

Therefore, we have shown that the Classical Physics with the Stochastic
Inertial-Gravitational Background gives the value of the Bell's observable
matching both the experimental data and the quantum mechanical value of the
Bell's observable. To sum it up, the description of microobjects by the
classical physics accounting for the effects brought about by the
Inertial-Gravitational Background is equivalent to the Quantum-Mechanical
descriptions, both agreeing with the experimental data.

\section{Conclusions}

We are regarding the inertial-gravitational background of isotopic fields
and waves as hidden variables. The inertial-gravitational background could
be considered negligible and not affecting the behaviour of quantum
microobjects. We are verifying whether this is correct. The quantitative
assessments of the inertial-gravitational background influence on the
quantum microobjects' behaviour have not been performed due to the former
having never been examined. The quantum effects are small as well, but their
quantitative limits are known and are determined by the Heisenberg
inequality. We have demonstrated the inertial-gravitational background being
random and isotropic to affect the phases of microobjects. Our case
corresponds to the Lagrange function $L(t,q,\dot{q},\ddot{q},...,\dot{q}%
^{(n)},...)$, depending on coordinates, velocities and higher time
derivatives, which we call additional variables, extra addends, or hidden
variables. In arbitrary reference systems (including non-inertial ones)
additional variables appear in the form of higher time derivatives
of coordinates, which complement both classical and quantum physics. We call
these additional variables constituting the higher time
derivatives of coordinates hidden variables 
complementing the description of particles. The contemporary physics
presupposes employment of predominantly inertial reference systems; however,
such a system is very hard to obtain, as there always exist external
perturbations, for example, gravitational forces, fields, or waves.
In this case, the relativity principle enables transfer from the
gravitational forces or waves to inertial forces. If the fact that the real
reference frames are non-inertial and hence there exist additional variables
in the form of inertial-gravitation effects is ignored, then non-local
correlation of quantum states and quantum non-locality would seem
surprising. The inertial-gravitational origin of quantum-mechanical wave
functions in the form of non-local hidden variables is described [7-9].


\end{document}